\documentclass[prb,superscriptaddress,longbibliography, twocolumn]{revtex4-1}

\usepackage{times}
\usepackage{amsmath}
\usepackage{epsfig}
\usepackage{color}
\usepackage{ulem}
\usepackage{graphicx}
\usepackage{dcolumn}
\usepackage{bm}
\usepackage{bookmark}
\usepackage{tabularx}
\usepackage{hyperref}
\usepackage{multirow}
\usepackage{lineno}
\usepackage{multirow}
\usepackage{tikz}
\hypersetup{colorlinks=true, citecolor=blue, filecolor=blue, linkcolor=blue, urlcolor=blue}

\begin{document}

\title{Atomistic Modeling of Martensitic Phase Transition in Hexamethylbenzene}

\author{Zarif Fahim}
\affiliation{Department of Mechanical Engineering and Engineering Science, University of North Carolina at Charlotte, Charlotte, NC 28223, USA}

\author{Pedro A. Santos-Florez}
\affiliation{Department of Mechanical Engineering and Engineering Science, University of North Carolina at Charlotte, Charlotte, NC 28223, USA}

\author{Qiang Zhu}
\email{qzhu8@uncc.edu}
\affiliation{Department of Mechanical Engineering and Engineering Science, University of North Carolina at Charlotte, Charlotte, NC 28223, USA}

\date{\today}

\begin{abstract}
Materials exhibiting a martensitic phase transition are essential for applications in shape memory alloys, actuators and sensors. Hexamethylbenzene (HMB) has long been considered as a classical example of ferroelastic organic crystals since Mnyukh's pioneering work in 1970s. However, the atomistic mechanism underlying this phase transition has never been clarified. In this work, we present a direct molecular dynamics simulation to investigate the phase transition mechanism in HMB. For the first time, we report a simulation results that can accurately reproduce both the transition temperature and hysteresis loop observed in previous experimental studies. By analyzing the MD trajectories, the potential energy surface, we identified that a low-barrier atomic sliding mode along the close-packed (11$\overline{1}$) plane of the low-temperature phase is the key to trigger the phase transition at the critical temperature window. This is further confirmed by the observed continuous softening of shear modulus around the transition window. Our results demonstrate that the integration of various atomistic modeling techniques can provide invaluable insight into the martensitic phase transition mechanisms in organic crystals and guide the development of new organic martensites.
\end{abstract}

\vskip 300 pt

\maketitle
\section{INTRODUCTION}
Phase transitions are critical in understanding the relationship between material structure and properties. Among various types of transitions, the diffusionless phase transition plays a pivotal role in numerous industrial applications, particularly shape memory alloys (SMAs) \cite{otsuka1999shape, otsuka2005physical}. When heat is applied to SMAs, they undergo a phase transition from low-temperature martensite to high-temperature austenite, a process known as the martensitic phase transition (MPT) \cite{mihalcz2001fundamental}. Many applications of SMAs (e.g. Ni-Ti alloy) are often incorporated into thin films, where they function as micro-actuators, making them highly valuable in various technological applications \cite{winzek2004recent, miyazaki1999martensitic, fu2004tini}. 

Despite extensive studies of polymorphic phase transition in metals, alloys or polymers \cite{oikawa2002magnetic, olsson2018ab, brown2007new}, MPT in molecular crystals have remained relatively underexplored. Molecular crystals have received growing attention in materials sciences due to their diverse group of properties ranged from thermal, mechanical to optoelectrical functionalities \cite{desiraju1995supramolecular, sutton2016noncovalent, park2018organic}. Unlike the inorganic counterparts, new molecular crystals with target properties can be more conveniently developed through modern crystal engineering using the rational design approaches \cite{desiraju1995supramolecular, sutton2016noncovalent, park2018organic}. 

Recently, the reconfigurability of molecular crystal has spurred the development of a burgeoning research field known as crystal adaptronics \cite{ahmed2018crystal}. Several organic crystals were experimentally reported to exhibit martensitic phase transitions \cite{takamizawa2019versatile, sasaki2019twinning, mutai2020superelastochromic, naumov2015mechanically, sahoo2013kinematic, panda2014colossal, colin2019thermosalient, khalil2019direct, diao2014understanding, chung2020understanding,su2015assembling, borbone2019high, das2010reversible}. The martensitic phase transitions in organic crystals have been found to couple with a range of functional properties, including thermoelasticity, thermosalience, superelasticity, and shape memory effects \cite{chung2020understanding}, which makes molecular crystals highly attractive for modern smart material applications, as they offer unique responsiveness to external stimuli and are well-suited for a variety of technological uses in areas like adaptive materials, sensors, and actuators.

Among molecular materials, hexamethylbenzene (HMB) is one of the very early examples that has been reported to exhibit the martensitic phase transition in the history of small molecule organic crystallography. At low temperatures, HMB molecules crystallize in a triclinic \textit{P}-1 symmetry (referred to as HMB-II) \cite{lonsdale1928}. In 1952, Saito studied the phase transition of HMB, and found that HMB-II undergoes a phase transition at high temperatures, forming what is now commonly referred to as the orthorhombic \textit{Fmmm} form I in the literature \cite{saito1952x}. Although the HMB II-I transition had been observed for some time, it was not initially recognized as an example of MPT at that time. In the 1970s, Mnyukh performed a series of pioneering studies on the polymophic phase transitions of molecular crystals \cite{mnyukh1972, mnyukh1973, mnyukh1975, mnyukh1976, mnyukh1977}. Among these studies, Mnyukh proposed that HMB, in together with several other systems, to possess a likely martensitic phase transition behavior \cite{mnyukh1975}. For HMB, he determined that form II transforms to I at 382-385 K upon heating and the reverse transition occurs at 380-382 K upon cooling. Based on X-ray data analysis, Mnyukh also suggested a molecular sliding mechanism to explain the crystallographic relation between two forms. Recently, Li and coworkers revisited the study of HMB’s phase transition and further confirmed its martensitic nature \cite{li2019martensitic}. More interestingly, they found that the rapid and complete structural switching behavior of HMB enables it to perform work comparable to that of many actuator materials. This exciting discovery suggests that organic crystals like HMB hold potential for applications in soft devices, such as actuators.


\begin{figure*}[ht]
    \centering
    \includegraphics[width=0.85\textwidth]{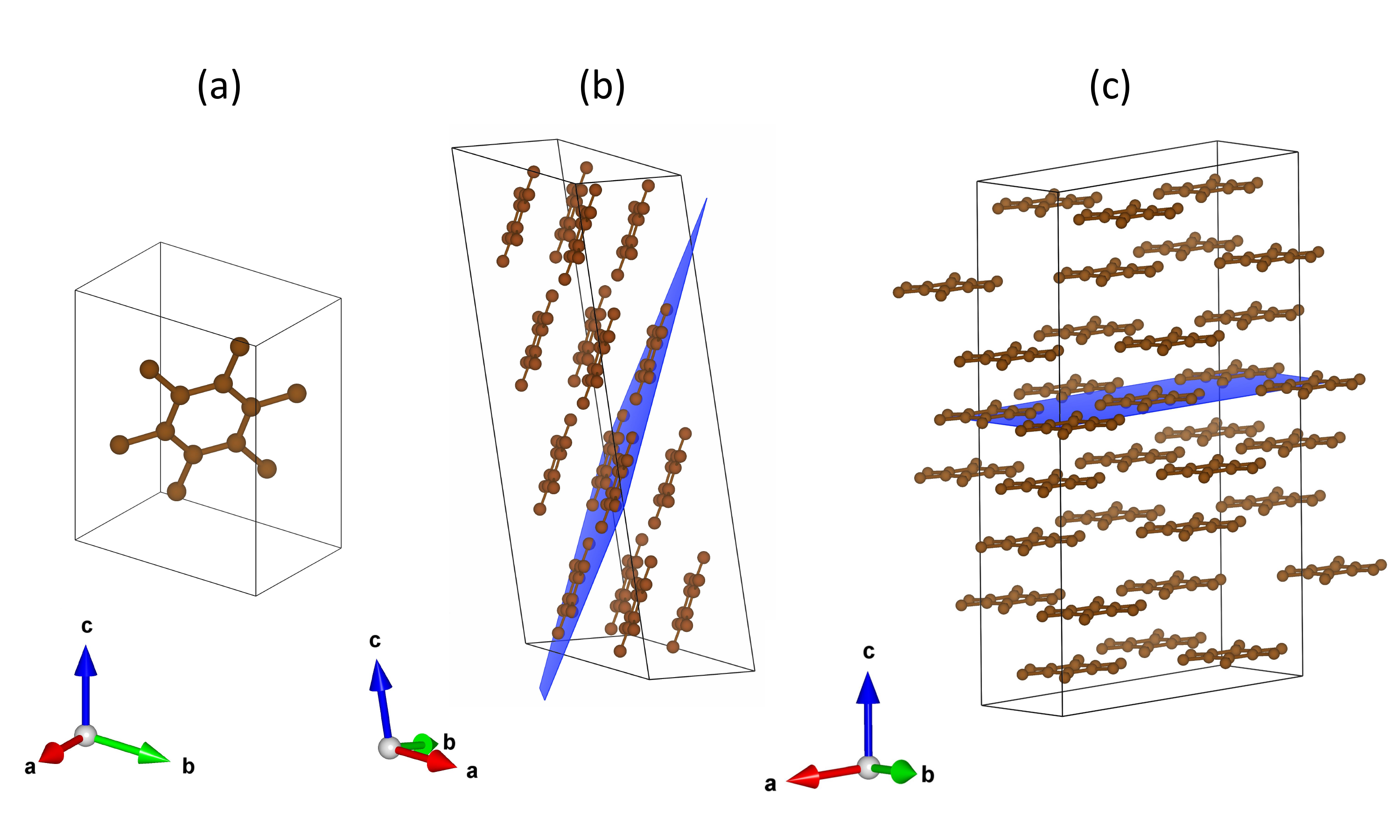}
    \caption{The crystal structure of HMB. (a) shows the HMB unit cell of with only one molecule; (b) displays the HMB supercell with the highlight of closest packing (11$\overline{1}$) plane (colored in blue) of the unit cell; (c) shows the rotated HMB supercell by placing the closest packing plane to the (001) plane (colored in blue) in the new setting.}
    \label{Fig1}
\end{figure*}

To our knowledge, identifying a new organic martensite in the laboratory often requires extensive trial and error through experimental efforts. Computational techniques, however, may offer the potential to accelerate this process by enabling high throughput \textit{in-silico} screening. To facilitate computational discovery, it is essential to first verify whether current modeling techniques can reliably reproduce experimentally observed phenomena. If successful, one may take steps further to develop a systematic computational pipeline to screen potential organic martensites from existing databases. 

In this work, we selected HMB as a model system to conduct molecular dynamics (MD) based atomistic simulation to elucidate the phase transition mechanism of HMB. In the following sections, we begin by providing the necessary computational details of our modeling approach. We then demonstrate how MD simulations can reproduce the experimentally observed MPT through simulated heating and cooling cycles of the HMB crystal. This is followed by an energy surface analysis to quantify low-energy barrier pathways during the phase transition. Additionally, we explore the use of elastic constants and stress-strain relations as indicators to evaluate the possibility of observing a MPT by starting from the known room-temperature phase. Finally, we conclude by discussing how our findings can facilitate the automated screening of new organic martensites.

\section{Computational Methods}

\subsection{HMB crystal and Model Setup}
As show in Fig. \ref{Fig1}a, the low temperature HMB form-II features a single molecule in the unit cell with \textit{P}-1 symmetry. Due to the molecular symmetry, the asymmetric unit contains a half molecule. In the standard crystallographic setting, it has a triclinic box with $a$= 5.260 \AA, $b$=6.199 \AA, $c$= 8.004 \AA, $\alpha$=103.8$^\circ$, $\beta$=98.7$^\circ$, $\gamma$= 100.2$^\circ$. Fig. \ref{Fig1}b displays a 2 $\times$ 2 $\times$ 4 super cell model based on the standard setting. Recently, we have implemented a molecular layer finding function into the open source Python library $\texttt{PyXtal}$ \cite{pyxtal} that can search for the molecular layer spacing in any arbitrary crystallographic plane. Using this module, we find that the HMB-II crystal has the largest separation of 3.595 ~\AA~ at the $(11\overline{1})$ plane (highlighted in blue) of the unit cell. This plane is also parallel to the benzene ring of the HMB molecule. As discussed in the previous section, it has been found that the shearing and molecular sliding activities on $(11\overline{1})$ plane is the main cause of phase transition. To simplify the analysis, we reoriented the unit cell model using a rotation matrix [1, 1, 2], [1, -1, 0], [4, 3, -1]. As depicted in Fig. \ref{Fig1}c, this transformation maps the $(11\overline{1})$ plane of the original unit cell onto the (001) plane of the rotated supercell. In addition, the resulting triclinic box has $\alpha$= 96.0$^\circ$, $\beta$= 91.0$^\circ$, $\gamma$= 90.0$^\circ$, which are closer to an orthorhombic box, thus allowing more robust neighbor finding in the MD simulation. Based on the rotated crystal, we created a 6$\times$5$\times$3 super cell model consisting of 345600 atoms in a triclinic box (see Table \ref{table1}) for the subsequent MD simulations.

\begin{table}[ht]
  \centering
  \caption{The cell parameters for \textit{P}-1 HMB-II phase at 293 K. }
  \label{table1}
\begin{tabular}{lrrrrrr}\hline\hline
Source~~~~~~~~~~~~~~~ &  ~~$a$~~ & ~~$b$~~ & ~~$c$~~ & ~~$\alpha$~~ & ~~$\beta$~~ & ~~$\gamma$~~ \\
&  (\AA) & (\AA) & (\AA) & ($^\circ$) & ($^\circ$) &  ($^\circ$)\\\hline
Unit cell-Expt.\cite{mnyukh1975} & 5.260 & 6.199 & 8.004 & 103.8 & 98.7 & 100.2   \\
Supercell-Expt.  & 185.083 & 88.112  & 173.514 & 96.0 & 91.0 &  90.0    \\
Unit cell-MD     & 5.372 & 6.442 & 8.257 & 105.1 & 100.1 & 98.1   \\
Supercell-MD    & 188.64 & 89.492 & 183.119 & 92.2 & 96.4  & 90.2  \\\hline\hline
\end{tabular}
\end{table}

\subsection{Molecular Dynamics Simulation}
It is well known that the accuracy of MD simulation results heavily depends on the quality of the interatomic force field. In this work, we employed our previously developed pipeline \cite{Santos-2023-PRR} to automate the generation of general amber force field (GAFF) interatomic potential \cite{case2021amber, wang2004development} using \texttt{AMBERTOOLS} \cite{case2021amber}. Specifically, we used the semiempirical AM1-BCC method \cite{jakalian2000fast} to compute the atomic partial charges for each atom in the molecule. To validate the force field, we performed NPT calculations of HMB form II at 300 K and obtained cell parameters that closely matched experimental values (see Table \ref{table1}) based on the standand settings. For the rotated supercell, the $a$, $b$ and $\gamma$ values match very well. However, the $c$ value is about 9.4\% mismatch. This is not unexpected since the layer spacing is more subtle due to the weak interlayer interactions. Due to the error in $c$, the variations on $\alpha$ and $\beta$ values are also notable. Despite the discrepancy in $c$, the overall agreement ensures the reliability of our subsequent simulation results. We used the \texttt{LAMMPS} \cite{LAMMPS} package to carry out the MD simulations, mimicking the experimental heating and cooling processes of the HMB crystal over a range of temperatures. In these simulations, we employed the Langevin thermostat \cite{adelman1976generalized} to control the temperature and the Nose-Hoover barostat \cite{hoover1985canonical} to maintain atmospheric pressure, with a 1 fs timestep. After testing several heating and cooling rates ranging from 3.33 $\times$ 10$^{9}$ to 1.25 $\times$ 10$^{10}$ K/s, we selected 4 $\times$ 10$^{9}$ K/s, which provided the optimal balance between accuracy and computational efficiency.

\subsection{Properties Evaluation}\label{methods}
To evaluate the mechanical and thermodynamic properties associated with MPT in our system, we conducted a comprehensive set of calculations, including $\gamma$-surface energy, elastic constants, and stress-strain relationships under mechanical loading. These analyses help us understand the structural and energetic factors influencing phase stability, deformation mechanisms, and the mechanical properties of HMB in this study.

In the context of stacking faults and plastic deformation in metals, the concept of the $\gamma$-surface energy has been widely used to evaluate the spatial distribution of penalty energy required for a crystal to glide along a particular plane. To compute the $\gamma$-surface energy, we employed a systematic approach where a chosen slip plane was displaced incrementally while relaxing the surrounding lattice. For each displacement, the total energy of the system was calculated using LAMMPS, capturing the energy landscape associated with partial or complete shifts of the crystal along the slip plane. This method allows us to map the energy variations across different slip vectors, thus building a comprehensive $\gamma$-surface profile. As we will discuss in Section \ref{gamma}, the resulting energy-surface profile is instrumental in predicting the ease of slip in the material, which is crucial to the martensitic transformation process.

For the computation of elastic constants, we start by applying Hooke’s law:
\begin{equation}\label{eq1}
\sigma_{ij} = C_{ijkl} \, \epsilon_{kl},    
\end{equation}

where $\sigma$ is stress tensor, $\epsilon$ is strain tensor, $C$ is the second order elastic constants, and $i, j, k, l$ are direction indices. Here, we employed LAMMPS to calculate the elastic constant of HMB. Specifically, we started by preparing a set of equilibrated structures at the desired temperatures by running the NPT MD simulation. For each equilibrated configuration, we applied a range of uniaxial and shear deformations in both positive and negative directions to measure the averaged stress values after an NPT equilibration with the constraints of strain deformation. The elastic constants were then derived from these stress values according to Eq. \ref{eq1}.

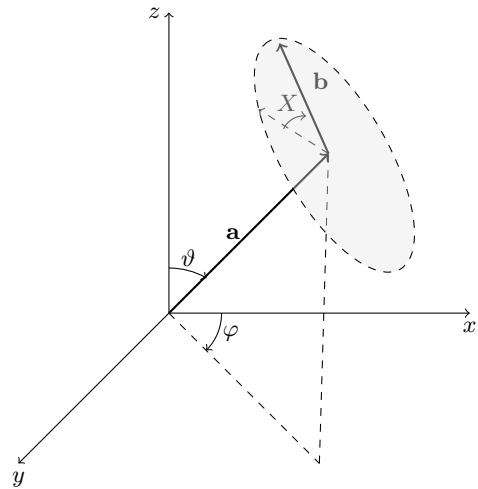
\begin{figure}[h!]
    \centering
    \begin{tikzpicture}
        \def\theta{90} 
        \def\phi{45}   
        \def\chi{45}   
        \def\delta{30}
        
        \coordinate (a) at ({3*cos(\phi)*sin(\theta)}, {3*sin(\phi)*sin(\theta)}, {3*cos(\theta)});
        
        \coordinate (b) at ({3*cos(\phi)*sin(\theta) + 1.5*cos(\theta)*cos(\phi)*cos(\chi) - 1.5*sin(\theta)*sin(\chi)},
                            {3*sin(\phi)*sin(\theta) + 1.5*cos(\theta)*sin(\phi)*cos(\chi) + 1.5*sin(\theta)*sin(\chi)},
                            {cos(\theta) - 1.5*sin(\theta)*cos(\chi)});
    
        \coordinate (c) at ({4.5*cos(\phi+\delta)*sin(\theta+\delta) + 0.01*cos(\theta+\delta)*cos(\phi)*cos(\chi+\delta) - 0.01*sin(\theta+\delta)*sin(\chi+\delta)},
                            {3*sin(\phi+\delta)*sin(\theta+\delta) + 0.01*cos(\theta+\delta)*sin(\phi+\delta)*cos(\chi+\delta) + 0.01*sin(\theta+\delta)*sin(\chi+\delta)},
                            {cos(\theta+\delta) - 0.01*sin(\theta+\delta)*cos(\chi+\delta)});
            
        \draw[->] (0,0) -- (4,0) node[anchor=north] {$x$};
        \draw[->] (0,0) -- (0,4) node[anchor=east] {$z$};
        \draw[->] (0,0) -- (-2,-2) node[anchor=north] {$y$};
    
        \draw[->, thick] (0,0) -- (a) node[midway, left] {$\mathbf{a}$};
    
        \draw[->, thick] (a) -- (b) node[midway, above right] {$\mathbf{b}$};
        
        \draw[->, dashed] (a) -- (c) node[midway, above right]{}; 
        
        \draw[dashed] (0,0) -- (2,-2) coordinate (a_proj);
        \draw[dashed] (a_proj) -- (a);
    
        \draw[->] (0,0.6) 
        arc[start angle=\theta, end angle=60, radius=1.0] 
        node[midway, anchor=east, xshift=6pt, yshift=5pt] {$\vartheta$};
    
        \draw[->] (0.7,0) 
        arc[start angle=0, end angle=-\phi, radius=0.7] 
        node[midway, anchor=south, xshift=5pt, yshift=-6pt] 
        {$\varphi$};
    
        \draw[<-] (a) ++(-0.3,0.5)
        arc[start angle=-275, end angle=-(250-\chi), radius=0.3] 
        node[midway, anchor=south, xshift=-2pt] {$X$};
    
        \draw[dashed, fill=gray!20, fill opacity=0.4,  rotate around={120:(2.2,2.1)}] (2.2,2.1) ellipse (1.75 and 0.7);
    
    \end{tikzpicture}
    \caption{The angular notation system \( \theta \), \( \varphi \), and \( \chi \) used to define the spatial shear modulus, where $\mathbf{a}$ is perpendicular to the shear plane (highlighted in gray eclipse), while $\mathbf{b}$ indicates the direction in which the strain is measured.}
    \label{Fig2:3d_vector_diagram}
\end{figure}

With the elastic constants determined, we calculated mechanical properties such as Young’s modulus and shear modulus and analyzed their spatial distribution in both 2D and 3D representations. For this analysis, we used the open-source Python library \texttt{ELATE} \cite{gaillac2016elate} to explore the anisotropic behavior of the shear modulus as a function of temperature. Since a shear mode depends on two orthogonal directions (the direction of applied shear stress and the direction in which shear strain is measured), we followed the notation system ($\theta, \phi, \chi$) as defined in \texttt{ELATE} to denote these shearing directions. For clarity, this notation is illustrated schematically in Fig. \ref{Fig2:3d_vector_diagram}, where vector $\mathbf{a}$ is perpendicular to the shear plane that measures the strain, while $\mathbf{b}$ indicates the direction in which the stress is applied.

To calculate the stress-strain relationship under mechanical loading, we adopted a similar approach to that used for elastic constants, but with a controlled, incremental application of strain. The deformation strain was applied at a specified rate (2$\times 10^{-7}$ fs$^{-1}$), and during the simulation, stress and strain values were recorded for further analysis. The simulation was run until a maximum strain of 20\% was reached, allowing us to capture the stress response and analyze the material’s deformation behavior under various loading conditions.

\section{Results and Discussions}

In the section, we discuss how to faithfully reproduce the experimentally observed phase transition, and extract the atomistic insight by analyzing direct MD simulations. Furthermore, we conducted a series of energy and property calculations to understand the origin of the martensitic phase transition from the perspective of atomistic interactions and molecular motions. Using HMB as a model system, we aim to explore the potential of integrating computational techniques to provide quantitative predictions of MPT in organic martensites.

\subsection{Heat Cycling Simulation}
We first attempted to set up the direct MD simulation to mimic the experimental heat/cool cycling as reported in a recent work \cite{li2019martensitic}. Specifically, we performed (i) an initial equilibration of 1 ns at 350 K under NPT ensemble; (ii) step-wise heating from 350 to 390 K over a period of 8 ns; (iii) a subsequent equilibration of 1 ns at 390 K; and (iv) reverse cooling from 390 K back to 350 K at the same rate used during the heating phase.

\begin{figure}[!htbp]
    \centering
\includegraphics[width=0.49\textwidth]{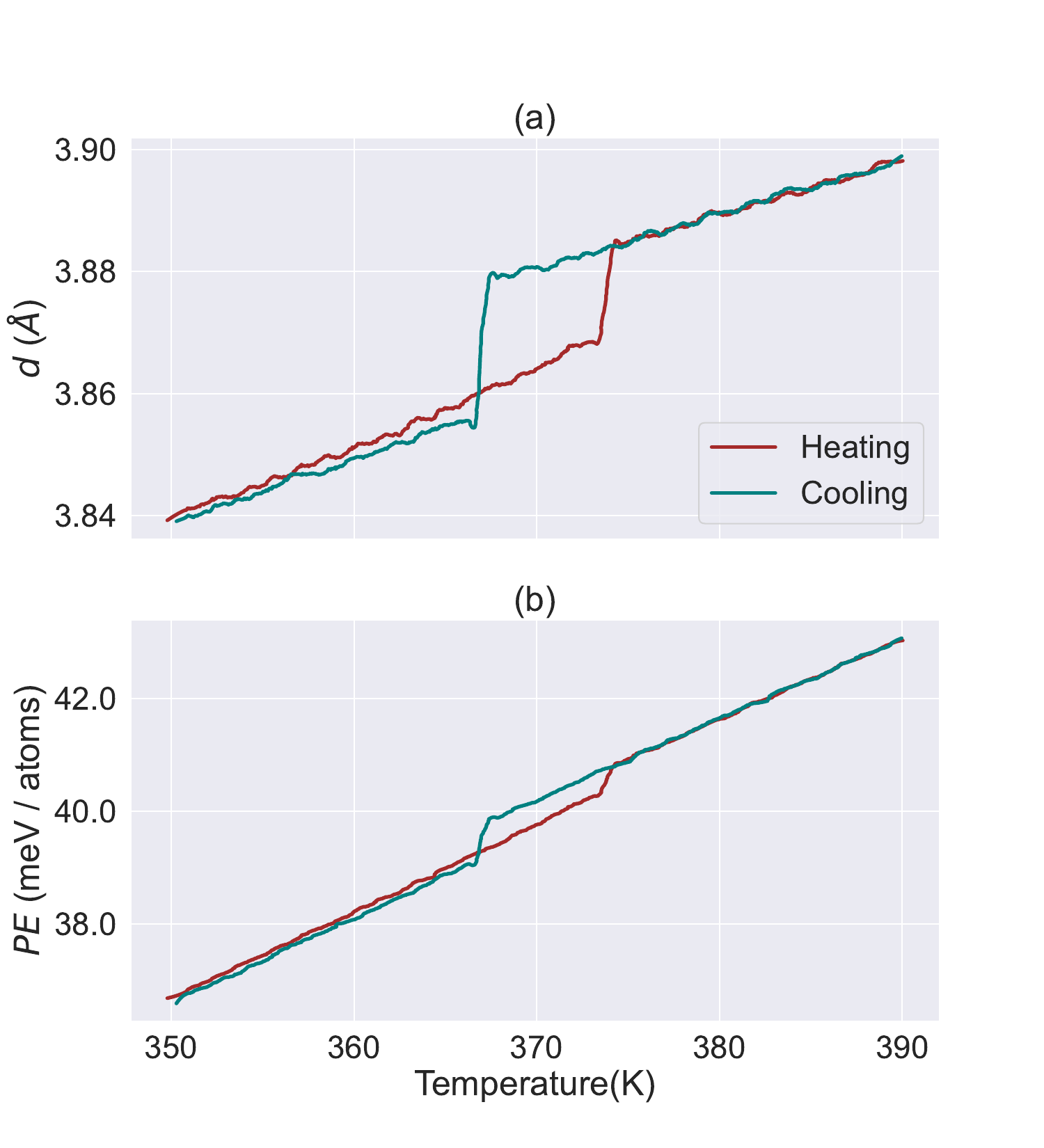}
    \caption{The evolution of two representative system variables upon the heating and cooling cycle simulation of HMB between 350 and 390 K. (a) shows the 
change of the averaged interlayer separation $d$ values along [001] against the temperature. (b) plots the averaged potential energy as a function of temperature.}
    \label{Fig3}
\end{figure}

Despite the significantly faster heating/cooling rate in our MD simulation  (4 $\times$ 10$^9$ K/s) compared to the practical experiments, our simulation successfully captures the phase transition behavior analogous to experimental observations. In Fig. \ref{Fig3}, we track the evolution of two representative variables—average interlayer spacing ($d$) and potential energy per molecule (\textit{PE}), as functions of temperature throughout the entire cycle. Both $d$ and \textit{PE}, taken as order parameters, clearly exhibit hysteresis loop behavior. During heating, we first observed a steady increase of $d$ up to 373 K, suggesting a normal thermal expansion behavior. Between 373 and 375 K, a sudden increase in $d$ within a short temperature window indicates the onset of a phase transition. Beyond 376 K, there is another steady expansion of $d$, indicating a thermal expansion of the new phase. A similar trend is observed during the cooling phase, except that the phase transition temperature window shifts to 366–368 K, resulting in a hysteresis loop in the overall curve. A similar hysteresis behavior is also observed in the evolution of \textit{PE} values, as shown in Fig. \ref{Fig3}b, further confirming the phase transition dynamics during the heating and cooling cycles.

\begin{figure*}
\centering
\includegraphics[width=0.9\textwidth]{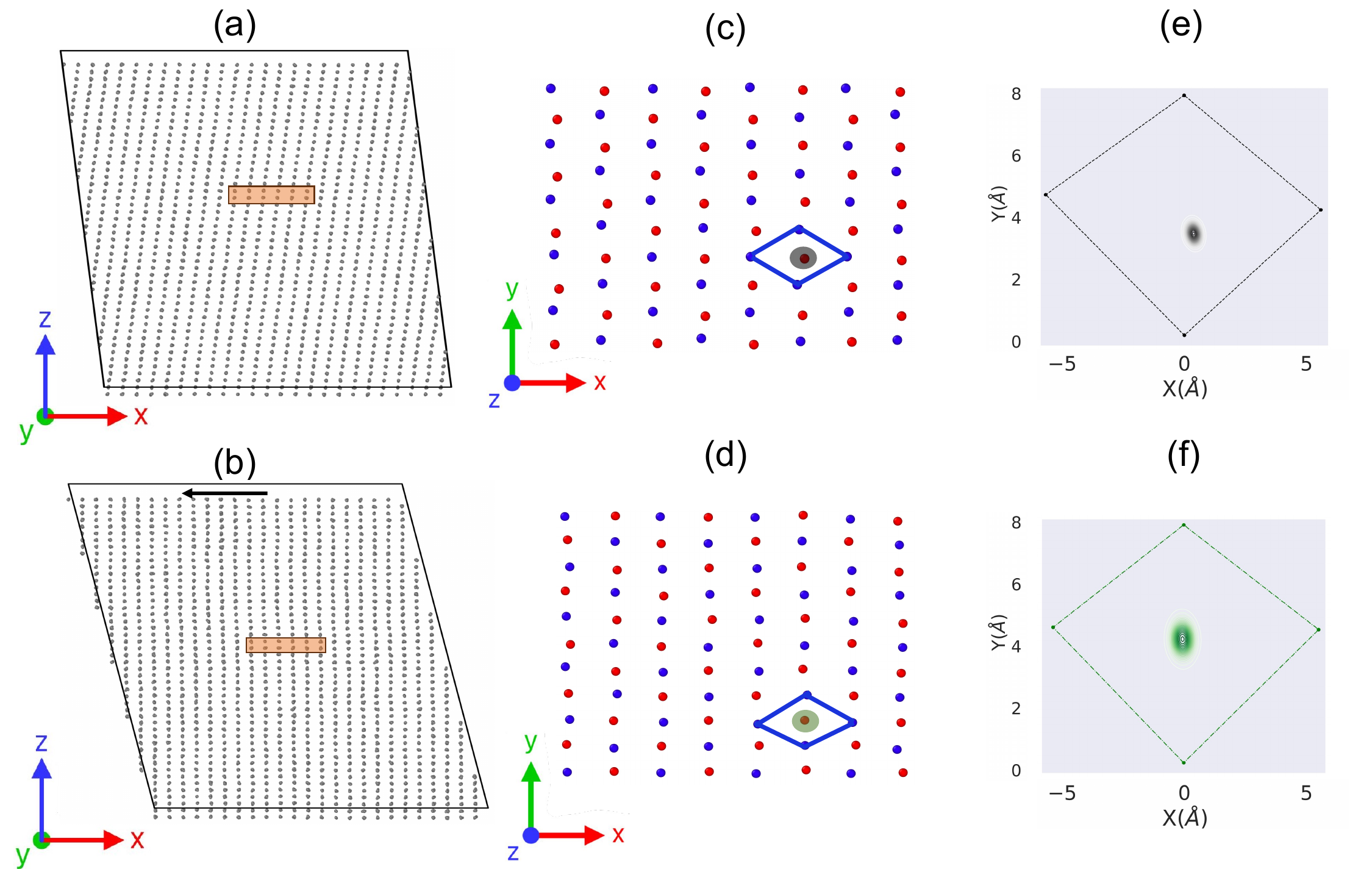}
\caption{Comparison of molecular arrangements in HMB before and after heat-induced phase transition. The upper panel depicts the low-temperature phase (HMB-II): (a) shows the side view in the $xz$ plane, (c) presents a zoomed top view of the rectangle region in (a) at the $xy$ plane, focusing on bilayer molecular centers colored in blue and red, and (e) illustrates the density distribution of each molecule relative to the nearest four molecular centers in the adjacent layer. The lower panel provides a similar analysis for the high-temperature phase (HMB-I) obtained from our MD simulation: (b) shows the side view, (d) presents the zoomed top view, and (f) shows the density distribution of molecular centers.}
\label{Fig4}
\end{figure*}

In previous experiments, Mnyukh \cite{mnyukh1975} and Li \cite{li2019martensitic} reported the phase transition happens around 382-385 K upon heating and 380-382 K upon cooling. Our simulations results are qualitatively consistent with the reported experimental values. We also attempted to model this process with different parameter settings. Indeed, we found that a slower cooling rate and a larger model size result in slightly higher phase transition temperatures. However, the overall physical behavior remains unchanged. Therefore, we conclude that our MD simulation can qualitatively reproduce the experimentally observed MPT phenomena, despite the limitations of force field accuracy and the faster cooling rates used in the simulations.

\subsection{MD Trajectory Analysis}

After confirming the reliability of our modeling approach, we proceed to study the atomistic mechanism of this phase transition by analyzing the molecular dynamics trajectories. Fig. \ref{Fig4}a-b plots two MD snapshots at 350 K and 390 K to represent the low temperature phase form II and high temperature phase form I, respectively. Clearly, we observe a rapid shearing on the $xz$ component during the transition window, corresponding to molecular sliding along the (001)[100] direction. 
Such a molecular sliding is expected to result in a change of stacking sequence between the adjacent (001) molecular layers. 

To better understand the sliding process, we select two adjacent (001) molecular layers and track the evolution of their molecular centers on the projected $xy$ plane as shown in Figs. \ref{Fig4}c-d. Notably, we found a common pattern in both phases. That is, each molecule (i.e, an individual red or blue sphere) is surrounded by four molecules from the neighboring layer that forms a rhombus like shape.
However, subtle differences in molecular positioning were observed between the phases. In the high-temperature phase, the shape is closer to an ideal rhombus, with the central molecule nearly located at the geometric center. In contrast, in the low-temperature phase, the shape deviates more from a perfect rhombus, and the central molecule is slightly displaced from the geometric center. 

We further calculated the distribution of red molecules relative to the blue rhombus by considering all MD trajectories during the phase transition. This results in a density plot as shown in Fig. \ref{Fig4}e-f. From the density plot, it can be seen that each HMB molecule in the high temperature phase is located at the center of the rhombus of the ajacent layer, while the HMB molecule in the low temperature phase is slightly off from the rhombus's center. These results suggest there is a systematic shift between the adjacent molecular layers upon heating. This process also results in a crystal symmetry change from triclinic \textit{P}-1 at the low temperature to high temperature orthorhombic \textit{Fmmm} symmetry \cite{mnyukh1975, saito1952x}.

Overall, the underlying molecular mechanism presented here closely aligns with Mnyukh’s interpretation, which was derived from experimental data \cite{mnyukh1975}. Additionally, we note that similar mechanisms have been observed in several other systems, as discussed in a recent review \cite{park2020martensitic}.

\begin{figure}[htbp]
\centering
\includegraphics[width=0.48 \textwidth]{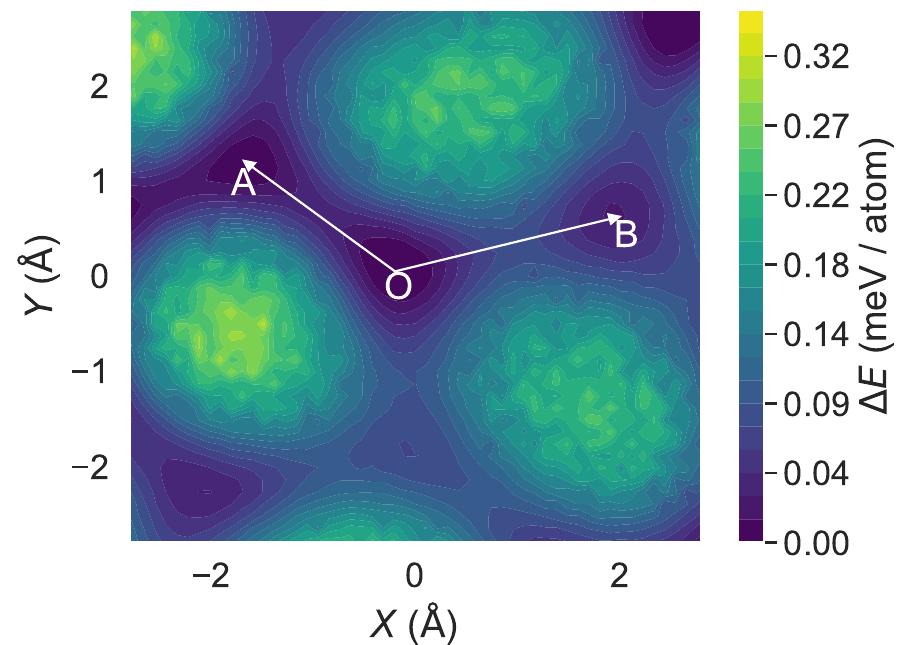}
\vspace{-5mm}
\caption{The calculated $\gamma$ surface energy contour plot for the rotated HMB supercell at the (001)-$xy$ plane.}
\label{Fig5} 
\vspace{-3mm}
\end{figure}

\subsection{$\gamma$-surface Energy Calculation}\label{gamma}
Although molecular sliding has been recognized as the key for the martensitic phase transition, the relationship between the sliding motion, crystal packing, and intermolecular interactions remains unclear. Following this approach, we performed the $\gamma$ surface calculation on HMB at 300 K. Specifically, we evenly split the super cell model into two regions along the $z$-axis, and then systematically displaced the upper region molecule in the entire $xy$ plane, while keeping the lower region fixed. For each displacement, the system was relaxed to the energy minimum under the rigid body assumption. Fig. \ref{Fig5} demonstrates the resulting energy surface contour plot that represents the difficulty of molecular sliding in terms of energy.

Starting from the origin \textbf{O} at (0, 0), we can see that there mainly exist two adjacent low energy basins (\textbf{A} and \textbf{B}). In particular, basin \textbf{A} has a slightly lower energy minimum value than \textbf{B}. In addition, the cost energy barrier from \textbf{O} to \textbf{A} is only 
0.082 meV / atom. Upon heating, the barrier is expected to become even lower. Such a low energy barrier value is comparable to magnitude of thermal fluctuations (23.6 meV at 300 K), thus the sliding from \textbf{O} to \textbf{A} can be easily triggered upon heating. 

We also found that the \textbf{O} $\rightarrow$ \textbf{A} direction (approximately [$\overline{1}10$] direction of the rotated cell) agrees with the molecular displacement in Figs. \ref{Fig4}f-g. The excellent agreements suggests that we may use the $\gamma$-surface energy analysis to predict the likely slip mode during the phase transition of molecular crystals without running long-time MD simulation. 

\begin{figure}[htbp]
    \centering    \includegraphics[width=0.49\textwidth]{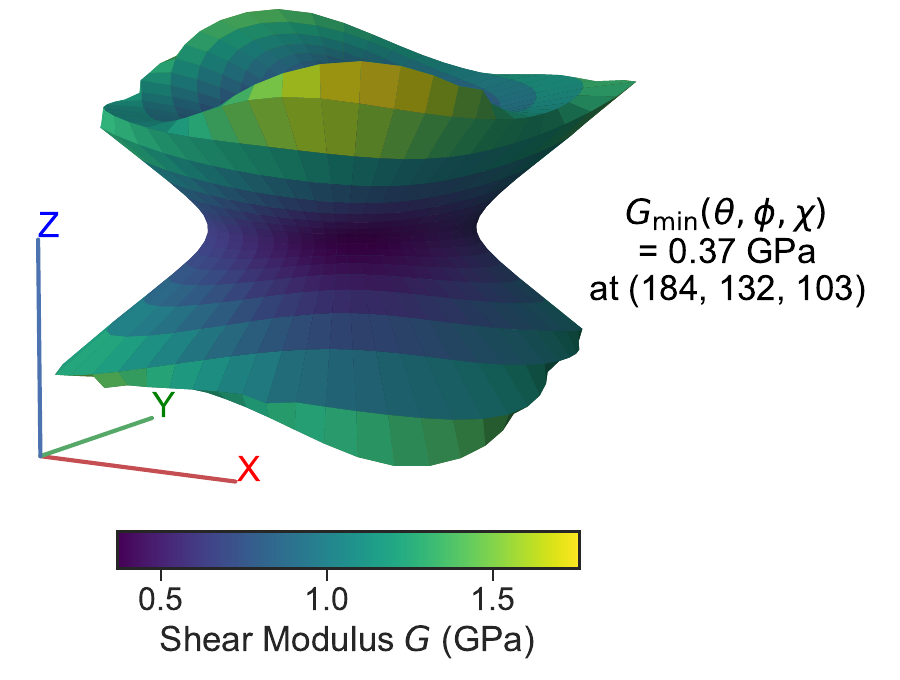}
    \caption{The calculated spatial distribution of minimum shear modulus values in HMB II at 300 K}
    \label{Fig6}
\end{figure}

\subsection{Evolution of Elastic Properties}

In addition to molecular sliding, the phase transition between HMB II and I can be alternatively understood from the mechanical shearing on the $xy$ plane. Hence, we investigated HMB's elastic properties as a function of temperature, following the method described in Section \ref{methods}.

We began with analyzing the three-dimensional plot of the minimum shear modulus for HMB-II at 300 K. As shown in Fig. \ref{Fig6}, the shearing plane ($\mathbf{a}$ direction) can be visualized imagining a vector connecting from the origin to any point on the plane. The distance from the origin to any point on a line represent the value of minimum shear modulus measurement among all $\mathbf{b}$ directions that is orthogonal to $\mathbf{a}$. Clearly, Fig. \ref{Fig6} suggests that the weakest shear modulus on the $xy$ plane are generally no more than 0.5 GPa and systematically smaller than other directions. In particular, the minimum shear modulus value of 0.37 GPa can be found at ($\theta=184, ^\circ$, $\phi=128 ^\circ$, $\chi=103 ^\circ$), which is approximately consistent (001) $xy$ plane and [$\overline{1}10$] direction as found in our previous $\gamma$-surface energy analysis.

\begin{figure}
    \centering
\includegraphics[width=0.5\textwidth]{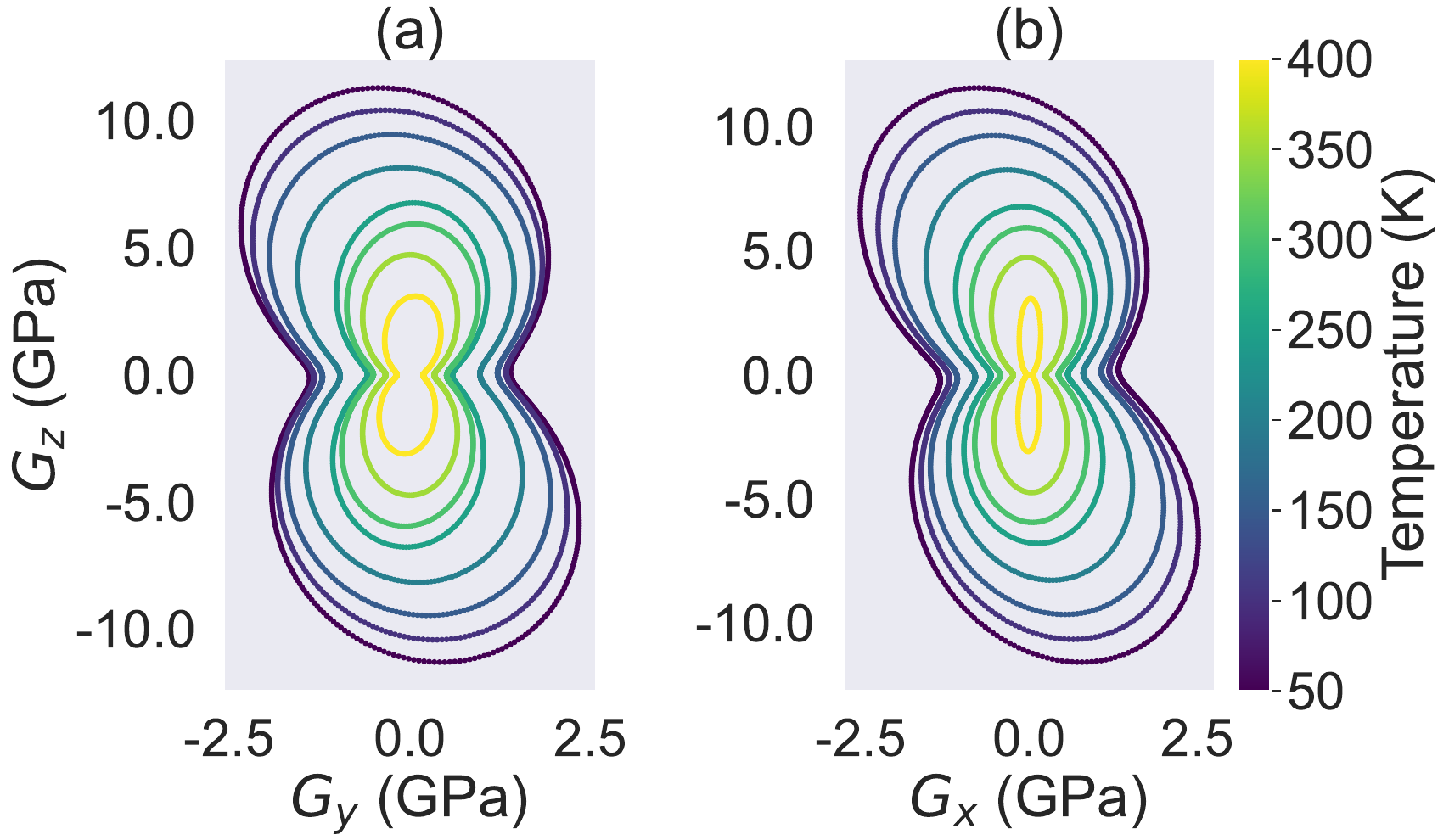}
    \caption{The evolution of HMB's shear modulus as a function of temperature at (a) $yz$ and (b) $xz$ planes.}
    \label{Fig7}
\end{figure}

Furthermore, we plotted the temperature-dependent evolution of shear modulus at $yz$ and $xz$ planes in Fig. \ref{Fig7}. In both plots, each contour line represents the shear modulus at a specific temperature. From Fig. \ref{Fig7}, it is evident that the shearing on $xy$ is much weaker than $yz$ and $xz$ planes. In addition, we can see that the shear modulus on all directions continuous decrease as the temperature increases, which is expected given that the thermal expansion should generally weaken the shear modulus.

\begin{figure}
    \centering
\includegraphics[width=0.45\textwidth]{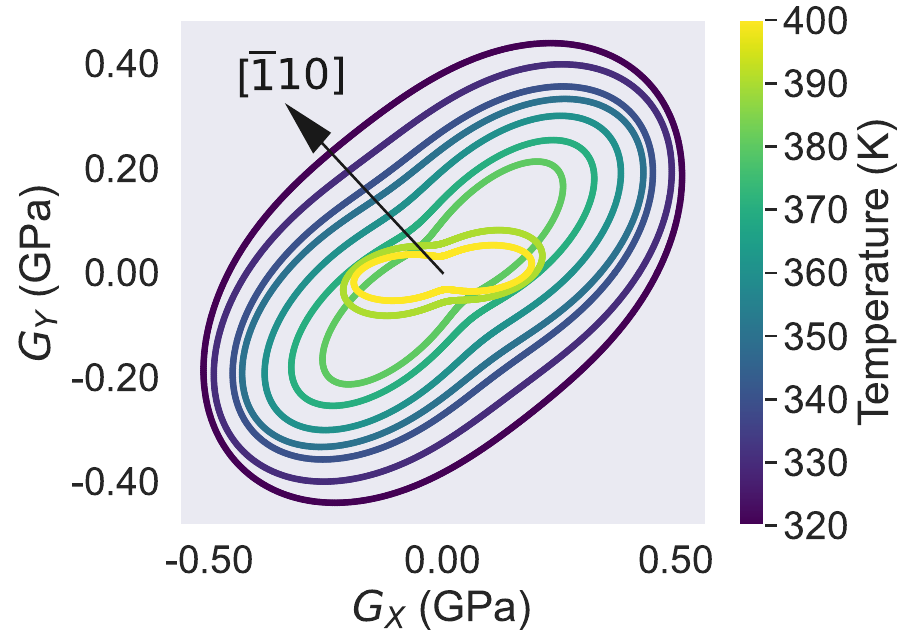}
    \caption{The evolution of HMB's shear modulus as a function of temperature at the $xy$ plane. The arrow denotes the direction leading to shear instability.}
    \label{Fig8}
\end{figure}

However, the continuous decreasing trend in shear modulus diminishes when we examine the evolution of shear modulus contour lines in the $xy$ plane. To illustrate this trend, we plot the contour line of the $G_{xy}$ with an interval of 10 K from 320 to 400 K, as shown in Fig. \ref{Fig8}. Between 320 and 380 K,  there is clearly a consistent decrease in shear modulus of HMB-II. To aid the analysis, a red line is drawn along the direction of minimum shear modulus on this plane, corresponding to the path from basin \textbf{O} to basin \textbf{A} as identified in our previous $\gamma$-surface calculation. Along this direction, the shear modulus at 380 K drops to approximately 0.1 GPa, indicating a substantial softening that suggests potential mechanical instability under shear. Indeed, further heating to 390 K produces a new contour shape, implying the onset of a new phase (HMB-I) due to shear instability. In HMB-I, the rearranged molecular stacking enhances mechanical stability by increasing the weakest shear modulus. These findings align well with our $\gamma$-surface analysis from the previous section, reinforcing the relationship between thermal effects and mechanical stability.


\subsection{The Impact of Mechanical Loads}
In addition to heat-induced phase transition, there has been a wealth of studies on molecular crystal phase transitions that can be triggered by mechanical work \cite{takamizawa2014superelastic, al2024ferroelastic, park2020martensitic}. Hence, we investigated the impacts of mechanical loading on HMB. In particular, we focus on the shear stress for a few representative temperatures and directions.

\begin{figure}
    \centering  \includegraphics[width=0.49\textwidth]{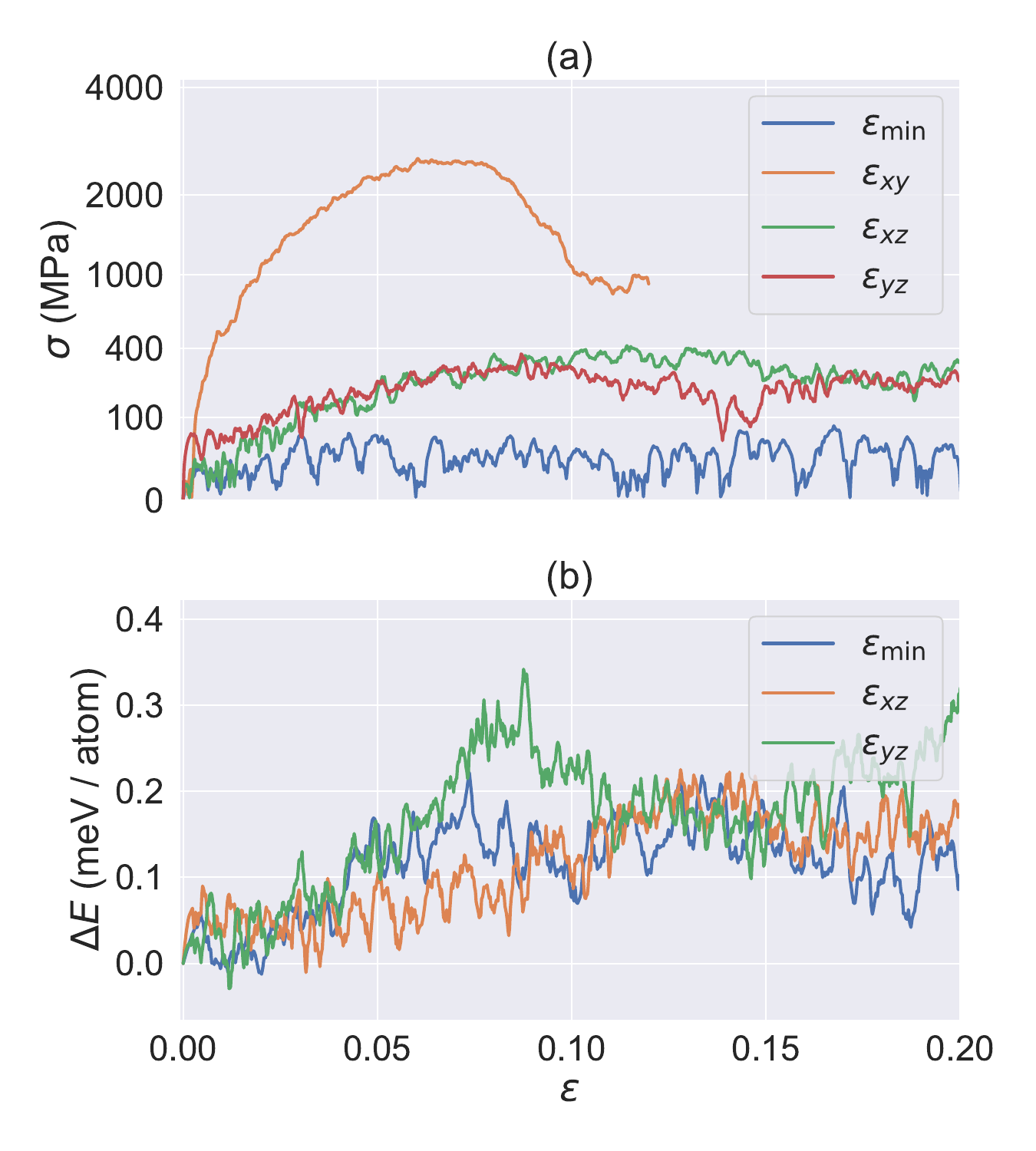}
    \vspace{-5mm}
    \caption{The simulation of HMB-II under various shear strains. (a) shows the stress-strain relation with the applied strain on $\epsilon_{xy}$, $\epsilon_{xz}$, $\epsilon_{yz}$ and the minimal stress $\epsilon_\text{min}$ directions at 300 K; (b) shows the corresponding energy change for $\epsilon_{xz}$, $\epsilon_{yz}$ and $\epsilon_\text{min}$. For clarity, all curves were smoothed by applying the Savitzky–Golay filter as implemented in \texttt{Scipy} with 20 coefficients.}
    \label{Fig9}
\end{figure}

Fig. \ref{Fig9}a plots the simulated shear stress-strain relation at three main directions $\epsilon_{xy}$, $\epsilon_{yz}$, $\epsilon_{xz}$, as well as the direction around [$\overline{1}10$](001) (will be called $\epsilon_\text{min}$ from now on) \cite{SM} for HMB I at 300 K. The results reveal that the peak shear stress in the $\sigma_{xy}$ direction (approximately 2500 MPa) is significantly higher than in the $\sigma_{xz}$ and $\sigma_{yz}$ directions (both around 400 MPa). Meanwhile, $\sigma_\text{min}$ exhibits a fluctuating stress response of approximately 50 MPa. These findings are consistent with our expectations, as the weakest interactions in the crystal are predominantly found between the $xy$ planes.

When the peak $\sigma_{xz}$ is reached at $\epsilon_{xz}=0.15$,  it corresponds to an approximate displacement of 0.54 \AA~ along the $x$-axis (assuming molecular spacing $d$=3.595 \AA). This displacement drives the structure away from the basin \textbf{O} towards to \textbf{B} as shown in our $\gamma$ surface energy plot (see Fig. \ref{Fig4}), thus leading to a phase transition and a subsequent decrease in the stress value $\sigma_{xz}$, as shown in Fig. \ref{Fig9}a. Under $\epsilon_{yz}$, the structure undergoes a phase transition from basin \textbf{O} to \textbf{A}. However, $\epsilon_{yz}$ does not align well with the \textbf{OA} direction. Hence it reaches a peak stress with less strain value ($\sim$0.1). In the case of $\epsilon_\text{min}$, the stress is even smaller, primarily due to the small shear modulus, as discussed in the previous section.

Fig. \ref{Fig9}b further compares the energy-strain relation for $\epsilon_{xz}$, $\epsilon_{yz}$ and $\epsilon_\text{min}$.
Since $\epsilon_{yz}$ does not align well with the \textbf{OA} direction, it requires a high penalty energy barrier in general. In contrast, both $\epsilon_{xz}$ and $\epsilon_\text{min}$ correspond to the direct deformation path along \textbf{OA} and \textbf{OB}, hence they require a lower energy barrier to cross. Indeed, we found $\epsilon_\text{min}$ is the preferred deformation path upon heating. On the other hand, the activation barrier of $\epsilon_{xz}$ is about the same as compared to $\epsilon_\text{min}$, suggesting that mechanical work of $\epsilon_{xz}$ could potentially trigger a phase transition. This may be interesting for the future experimental studies. 

Finally, the barrier values are about $\sim$0.30 meV/atom for $\epsilon_{yz}$, and $\sim$0.20 meV/atom for $\epsilon_{xz}$ and $\epsilon_\text{min}$. These values are about 3 times larger than the barriers found in $\gamma$-surface calculation. This discrepancy is expected, as the $\gamma$-surface setup assumes a collective slip of one entire plane, while shear deformation involves the continuous slipping of multiple planes.

\section{Concluding Remarks and Outlook}
In this work, we present a direct molecular dynamics simulation to investigate the phase transition mechanism in the model system HMB. Our results demonstrate that direct MD simulation of heating and cooling can accurately reproduce both the transition temperature and hysteresis loop observed in previous experimental studies.

Through analysis of the MD trajectories, we found that a molecular slip mode along (001)[$\overline{1}10$] in the rotated supercell is the key mechanism triggering the phase transition within a critical temperature range. Further $\gamma$-surface energy analysis confirms that this mode has a low energy barrier that can be activated by thermal vibration around 380 K. This result is also supported by the observed continuous softening of shear modulus at (001) plane, in particular along the [$\overline{1}10$] direction. Collectively, these results highlight that the integration of different atomistic modeling techniques can provide invaluable insights into the mechanisms underlying martensitic phase transitions in organic crystals.

For the case of HMB, it is evident that the crystal packing with large molecular layer separation allowing for slip or shear at low activation barriers, is fundamental to enabling this martensitic phase transition. Indeed, this class of molecular crystals has been widely recognized for its exceptional mechanical properties \cite{reddy2005structural}.

The computational pipeline developed through this work, from molecular plane analysis and MD simulation to property calculation, is entirely general for the study of other systems which may exhibit similar structural behaviors. Our studies on HMB, along with recent works \cite{Santos-2023-PRR, al2024ferroelastic}, demonstrate that generic force fields such as GAFF are sufficient for qualitative simulations of phase cycling. Additionally, atomistic simulations offer quantitative insights—such as energy barriers and elastic properties—that help predict the feasibility of phase transitions driven by external stimuli, such as heat or mechanical work. Looking ahead, we aim to automate this pipeline to streamline data collection and enable comparative analyses across a wide range of organic materials, ultimately guiding the design of mechanically flexible organic systems.

\begin{acknowledgements}
This research is sponsored by the NSF (DMR-2142570). The computing resources are provided by ACCESS (TG-MAT230046). The authors also thank Pance Naumov and Liang Li for helpful discussions.
\end{acknowledgements}


\bibliography{ref}

\end{document}